# Comment on "Feasibility of an electron-based crystalline undulator"

V.M. Biryukov[♦]

*Institute for High Energy Physics, Protvino, 142281, Russia*

**Abstract**

Tabrizi *et al*. [1] discuss the feasibility of an electron-based crystal undulator (e-CU) by planar channeling of 50 GeV electrons through a periodically bent crystal. We show that their scheme is not feasible. First, their undulator parameter is $K \gg 1$ always, which destroys photon interference. Second, they overestimate the electron dechanneling length $L_D$ in e-CU by an order of magnitude, which shortens the number N of e-CU periods from 5-15 (as they hope) to just 1-2. This kills their e-CU concept again. We made first simulation of electron channeling in undulated crystal and conclude that an electron-based crystal wiggler is feasible with wiggler strength $K \approx 10$ and number of periods $N \approx 2$.

In undulators, parameter *K* is the ratio of the maximal trajectory angle of a particle $\theta_{MAX}$ to the natural opening angle of the synchrotron radiation cone of $1/\gamma$, i.e. $K = \gamma \theta_{MAX}$ or $\theta_{MAX} = K/\gamma$. For very large *K* the opening angle (in the undulation plane) of the radiation is correspondingly large and no interference of the photons takes place. For very small *K* the radiation from all undulator periods overlaps and strong interference effects occur. In ref. [1] Tabrizi *et al*. correctly notice that "moderate values of the undulator parameter, $K \sim 1$, ensure that the emitted radiation is of the undulator type". Despite this requirement, Tabrizi *et al*. didn't give the undulator parameter for their schemes; however, it happens to be $K \gg 1$ in every scheme!

In magnet undulators, all particles move in the same field and thus have the same *K*. In crystal undulators, particles have channeling oscillations w.r.t the crystal planes which in turn undulate with a given period; see references in [5,6]. Thus, the particle trajectory angle w.r.t. the lab frame is defined by the sum of two motions, channeling and undulation.

The 50-GeV electrons channeled in unbent Si(111) oscillate w.r.t. the crystal planes with an angle aperture of about Lindhard angle $\theta_L = 32$ μrad, which already significantly exceeds the radiation emission angle of $1/\gamma = 10$ μrad. Oscillations with $\pm \theta_L$ set an undulator parameter of

---
[♦] http://mail.ihep.ru/~biryukov/

$K=\gamma\theta_L=3.2$. Further on, Tabrizi *et al.* require the Si(111) planes to undulate with an amplitude (much) bigger than the channeling oscillation amplitude. For the e-CU schemes [1], the maximal trajectory angle of an electron channeled in e-CU corresponds to $K=\gamma\theta_{MAX}=6\text{-}13$. These values, $K\gg1$, make the device a wiggler, not undulator. The difference is not semantic.

In undulator, $K\leq1$, the radiation from all undulator periods overlaps and strong interference effects occur. The angular aperture of the radiation (from a single electron) is narrowed down, because of interference, to $\theta_U\approx 1/(\gamma N^{1/2})$ so that for large N the angular spectral density of radiation (or brightness) can become extremely high and it grows as $N^2$.

In a wiggler with $K\gg1$ the radiation aperture angle (in the bending plane) is $\theta_W\approx K/\gamma$, i.e. much larger than the 'natural' aperture of the radiation. Interference of photons is suppressed. The brightness of a wiggler is orders of magnitude less than that of an undulator and grows just linearly with N.

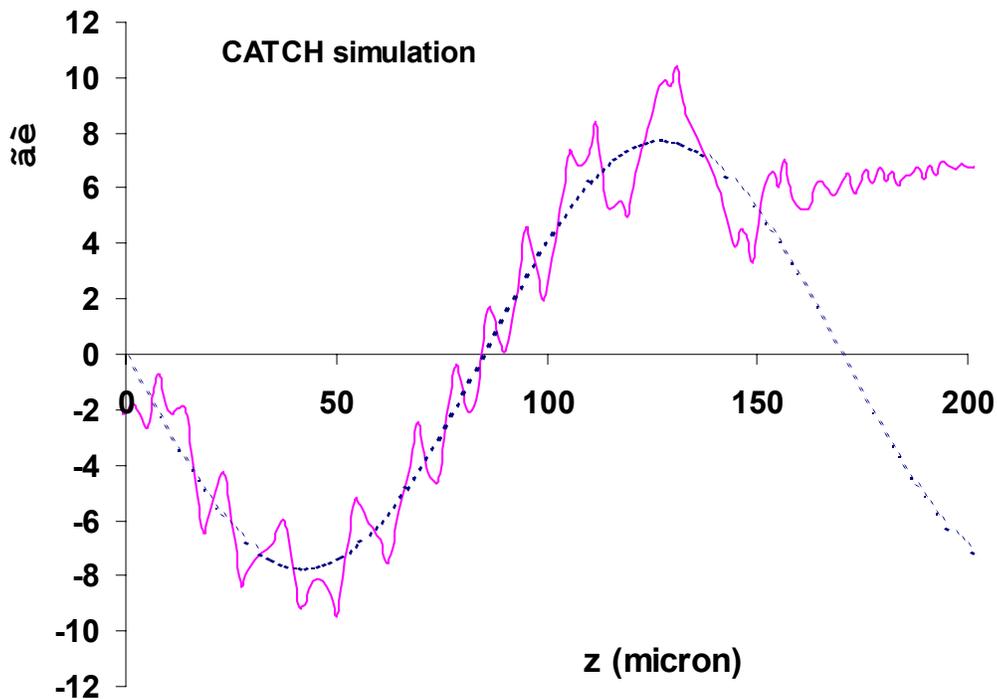

**Figure 1** Electron trajectory angle $\theta$ plotted (solid) as $\gamma\theta$ (z) along a Si (111) crystal undulator, CATCH simulation. Si(111) plane angle (dashed).

Figure 1 plots $\gamma\theta$ for a realistic trajectory angle $\theta$ of an electron channeled in one of e-CU (period $\lambda=0.17$ mm, N=5 [1]) simulated by CATCH code [2]. Here, $K\approx10$ as seen in Fig. 1. Electron (solid line) follows the Si(111) planes (dashed) over ~1 period of CU and then dechannels. Our simulations find that dechanneling length $L_D$ (hence, the number of periods $N\approx L_D/\lambda$) may be strongly overestimated in ref. [1].

Dechanneling length $L_D$ is another key issue of e-CU as it defines the length (and thus the achievable number of periods N) of an undulator. Tabrizi *et al.* assume $L_D \approx 1.2$ mm in unbent Si(111) planes, defined simply as the length of multiple scattering in amorphous Si at angle $\theta_L$. This elementary estimate for kinetic process might give the right order of magnitude provided it is justified by experiments or accurate computation based on kinetic theory or simulations. Of course, to know the number of undulator periods just by the order of magnitude is not enough for a feasibility proof.

Numerical solution of Fokker-Planck equation gave $L_D$=13 µm/GeV for electron in Si(110) [3] i.e. 0.65 mm at 50 GeV or factor of 2 lower than Tabrizi *et al.* estimate. Our CATCH simulation gives $L_D$=0.33 mm in unbent Si(111) at 50 GeV or factor of 3.6 lower than Tabrizi *et al.* estimate. Few experiments measured electron $L_D$ in Si(110) planes but gave the same $L_D$= 28-36 µm for any energy from 54 to 350 to 1200 MeV [3]. These sparse data do not support the conjecture of $L_D \approx 1.2$ mm in Si(111) at 50 GeV. Note that in Si(111) an electron is channeled by double plane, unlike in (110).

Further on, Tabrizi *et al.* estimate $L_D$ in a crystal bent with radius R in harmonic approximation $L_D(C)=L_D(0)(1-C)^2$, here $C=R_C/R$ and $R_C$ is critical radius. Harmonic approximation is a simplified result of the diffusion model for *positive* projectiles [4]. Namely, the channeled positive projectiles move far from crystal nuclei, so the diffusion coefficient for positives in Silicon is nearly a linear function of the transverse energy—*this is* why diffusion model predicts a proportionality of $L_D$ to the critical transverse energy which leads to harmonic approximation for $L_D$ [4]. For *negative* particles these assumptions are just wrong and harmonic approximation is nonsense. To prove their point, Tabrizi *et al.* would have to find diffusion coefficients for negative projectiles then solve the kinetic equation for electron dechanneling analytically and show it can be reduced to harmonic (or not) approximation, but they didn't come to this issue.

For C=0.2, they have $(1-C)^2$=0.64 while CATCH simulation finds for electrons $L_D(C)$= 0.33$L_D$(0) in Si(111). Forster *et al.* [4] measured 0.41 for protons in Si(110) and C=0.2, i.e. even for positives the harmonic approximation is overly optimistic for strong bending. As an overall result, CATCH gives $L_D$=0.11 mm at 50 GeV for electrons with C=0.2 in Si(111) while Tabrizi *et al.* use $L_D$=0.85 mm. Our evaluation of $L_D$ shortens the number of periods realizable in e-CU schemes [1] to just 1-2 instead of 5-15. This ruins e-CUs of ref. [1] again and strongly cuts e-CU brightness yet further. We summarize that an undulator requires $K \leq 1$ and $N \gg 1$. Instead, we have $K \gg 1$ and $N \approx 1-2$, a wiggler.

Tabrizi *et al.* do not mention that CUs were already produced [5] and their radiation in positron beam studied [6].